\documentclass[preprint]{elsarticle}

\usepackage{lineno,hyperref}
\modulolinenumbers[5]
\usepackage{dsfont}
\usepackage{amsmath}
\usepackage{braket}
\usepackage{amsfonts}
\usepackage{amssymb}
\usepackage{siunitx}
\usepackage{graphicx,subfigure}  
\usepackage[dvipsnames]{xcolor}
\usepackage{todonotes}
\usepackage{easyReview}
\usepackage{bbm}
\usepackage{physics}
\usepackage{mathtools}
\usepackage{ulem}
\journal{arXiv}
\date{}

\begin{document}

\begin{frontmatter}

\title{Quantum theory of statistical radiation pressure in free space}

\author[1]{Navdeep Arya\corref{cor1}%
	\fnref{fn1}}
\ead{navdeeparya.me@gmail.com}
\author[2]{Navketan Batra}
\ead{navketan_batra@brown.edu}
\author[1]{Kinjalk Lochan}
\ead{kinjalk@iisermohali.ac.in}
\author[1]{Sandeep K. Goyal}
\ead{skgoyal@iisermohali.ac.in}

\cortext[cor1]{Corresponding author}
\fntext[fn1]{First author}
\address[1]{Department of Physical Sciences, Indian Institute of Science Education and Research (IISER) Mohali, Sector 81 SAS Nagar, Manauli PO 140306 Punjab India.}
\address[2]{Brown Theoretical Physics Center and Department of Physics, Brown University, Providence, Rhode Island 02912, USA.}
\begin{abstract}
Light is known to exert radiation pressure on any surface it is incident upon, via the transfer of momentum from the light to the surface. In general, this force is assumed to be pushing or repulsive in nature. In this paper, we present a complete quantum treatment of radiation pressure. We show that the interaction of an atom with light can lead to both repulsive and attractive forces due to the absorption and emission of photons, respectively. An atom prepared in the excited state initially will experience a pulling force when interacting with light. On the other hand, if the atom is prepared in the ground state then the force will be repulsive while having the same magnitude as in the earlier case. Therefore, for an ensemble of atoms, the direction of the net force will be decided by the excited and ground state populations. In the semi-classical treatment of light-matter interaction, the absorption and emission processes have the same probability. Therefore the magnitudes of the force in the two processes turn out to be the same. We obtain the effective emission profile for an excited atom interacting with a quantum electromagnetic field, and show that in the quantum treatment, despite these probabilities being different, the magnitudes of the two statistical forces remain the same. This can be explained by noting that the extra contribution in the emission process is due to the interaction of the atom with the vacuum modes of the electromagnetic field, which results in a symmetric emission profile, contributing to a net zero force on the atoms in an ensemble. We further identify the set of states of electromagnetic field which give rise to non-zero momentum transfer to the atom.  
\end{abstract}


\end{frontmatter}

\linenumbers

\section{Introduction}
In the analysis of light matter interaction, it is well known that light exerts a pushing force through the radiation pressure on matter particles that it interacts with. This resulting force  can be used to accelerate, trap and sort matter particles~\cite{ashkin1970acceleration,ashkin1986observation}. Interestingly, there can be situations where the light exerts a pulling force, rather than pushing one, on an object, a phenomenon known as the {\it negative radiation pressure}. The possibility of negative radiation pressure was first mentioned in~\cite{Veselago_1968} and subsequently there have been considerable efforts to achieve it, for example, by simultaneously using two oppositely directed beams, using two beams with different longitudinal wave numbers, using gain mediums, and using a vector Bessel beam~\cite{shvedov2010giant, sukhov2010concept, mizrahi2010negative, chen2011optical, Forgacs2013}. In such studies, various  properties of classical light in the presence of dielectric media play an important role. 

The quantum character of electromagnetic field endows it  with quantum fluctuations which play central role in some of the most fundamental phenomena in nature, which include the Casimir effect, the Unruh effect, the Hawking radiation among others~\cite{casimir1948influence,unruh1976notes,hawking1974black}. Two of the well known phenomena in the context of atomic physics, which are sensitive to the states of the electromagnetic field and hence the quantum fluctuations therein,  are the spontaneous and the stimulated emissions from atoms.
By utilizing the spontaneous emission from the atom some of the most prominent predictions in quantum field theory can be tested in laboratory settings~\cite{lochan2020detecting}. It is further argued that spontaneous emission can cause friction-like effect in  an inertially  moving atom~\cite{sonnleitner2017will}. On the other hand, stimulated emission is known to exert optical forces in bichromatic and polychromatic techniques~\cite{voitsekhovich1991stimulated,PhysRevLett.71.3087,chieda2011prospects,jayich2014continuous,romanenko2014cooling,dai2015efficient,aldridge2016simulations,yang2016bichromatic,kozyryev2018coherent,cashen2003optical}. 

In this paper, we present a complete quantum treatment of radiation pressure. We consider the interaction of an atom with quantized electromagnetic field and show that the initial state of the atom plays a vital role in deciding the sign of the radiation pressure evaluated statistically. There are quantum states where the emission or absorption statistically do not lead to any momentum transfer, but one can also identify class of states which do lead to a non-zero momentum transfer on the average.

For an atom, the stimulated emission  results in a force opposite to the direction of the emitted photon, i.e, negative radiation pressure, due to the recoil, whereas the absorption results in a pushing force in the direction of the absorbed photon. Therefore, for an ensemble of atoms, whether the emission or absorption process is dominant will decide the attractive/repulsive character of the net force.

We illustrate that certain classes of quantum states, e.g., optical coherent state and Fock states  of electromagnetic field result in non-symmetric emission profile under dipole mediated quantum emissions, which do not cancel out over the full angular range,  leading to a non-zero net force. On the other hand, field states leading to symmetric emission, such as thermal radiations, do not cause any net acceleration. Therefore, in vacuum state of the electromagnetic field, there is no velocity change of the atom. Thus, the emission profile provides a straightforward  way of identifying quantum states of the field leading to zero and non-zero accelerations both~\cite{sonnleitner2017will,sonnleitner2018mass}.

Additionally, the study of atomic transition probabilities and the momentum transfer to the atom due to its interaction with pulsed Fock and coherent states of light is useful as efficient energy transfer between the atomic and radiation degrees of freedom has wide applications in the realization of quantum networks~\cite{Kimble2008,Duan2010,Stobiska2009,Wang2011,Rag2017}.

\section{Emission profile and momentum transfer}\label{Sec:Emission profile and momentum transfer}
Let us consider a two-level atom with  $\{\ket{g},\ket{e}\}$ representing its ground and excited states, and the free Hamiltonian $H_s = \hbar\omega_0\ket{e}\bra{e}$.  The atom is also interacting with electromagnetic field, through the interaction Hamiltonian~\cite{takagi1986vacuum}  given by $ H_{I} (\tau) \equiv d^{\mu}E_{\mu}  =  - {{\bf\mathcal{D}}}(\tau) \cdot {\mathbf{E}}(\tau)$ in the rest frame of the atom, identified by its proper time $\tau$.We will use the interaction Hamiltonian in the rotating-wave approximation. Here
\begin{align}
	{\bf E}(\tau) &=  \sqrt{\frac{\hbar}{\epsilon_0}} \int \frac{d^3\vb{k}~\text{i}\omega_k}{\sqrt{2 \omega_{k} (2\pi)^3}}  \sum_{\lambda_{\bf k}=1}^{2} {\va{\epsilon}}_{{\bf k}\lambda_{\bf k}} \left(a_{{\bf k}\lambda_{\bf k}} e^{-\text{i} \omega_k \tau} - H.c.\right), \label{Eq:field}
\end{align}
and 
${\mathcal{D}}(\tau) = \mathbf{d} ~ \sigma_{-} e^{-i \omega_{0} \tau} + \mathbf{d}^* ~ \sigma_{+} e^{+ i \omega_{0} \tau}$,
are the electric field and electric dipole moment operators, respectively, $\mathbf{d} = \bra{g}{\mathcal{D}}(\tau=0)\ket{e}$ is the transition dipole moment which we assume to be non-zero only along the $z$-axis, and ${a}_{{\bf k}\lambda_{\bf k}}~({a}^{\dagger}_{{\bf k}\lambda_{\bf k}})$ annihilates (creates) an excitation in the mode $({\bf k}, \lambda_{\bf k})$,  having polarization denoted by $\lambda_{\bf k}$, with $\sigma_+ = \sigma_-^\dagger = \ket{e}\bra{g}$ being the step up operator for the atom. The ${\va{\epsilon}}_{{\bf k}\lambda_{\bf k}}$ are the polarization unit vectors and can be chosen to be $\va{\epsilon}_{\mathbf{k},1} = (\cos \theta_{\mathbf{k}} \cos \phi_{\mathbf{k}}, \cos \theta_{\mathbf{k}} \sin \phi_{\mathbf{k}}, - \sin \theta_{\mathbf{k}}) ~~ \text{and,} ~~ \va{\epsilon}_{\mathbf{k},2} = (- \sin \phi_{\mathbf{k}}, \cos \phi_{\mathbf{k}}, 0).$

If the atom is prepared initially in the excited state $\ket{e}$, and the field is in a state $\ket{\Psi}$, then the final state $\ket{\Psi^e_f}$ of the atom plus field reads
\begin{align}
	\ket{\Psi^e_f} = \mathcal{T} \exp\left(-\frac{i}{\hbar}\int_{-\infty}^{\infty} \dd{\tau} H_I(\tau)\right) \ket{\Psi^{e}_i},
\end{align}
where $\ket{\Psi^{e}_i} = \ket{e}\otimes \Ket{\Psi}$ is the initial state of the composite system and $\mathcal{T}$ is the time-ordering operator. 
We can define the momentum operator, $P^{\mu} \equiv (P^0,\vb{P})$, for the field as
\begin{align}
	P^{\mu} &= \hbar\sum_{\lambda=1}^{2} \int {\rm d}^3k ~k^{\mu} a_{\vb{k}\lambda}^\dagger a_{\vb{k}\lambda},
\end{align}
where $k^{\mu} = (\omega_k/c, \vb{k})$. Since the field and the atom constitute a closed system, if we calculate the momentum in the initial and final states of the field, then the difference between the two must be the momentum transferred to the atom, due to the conservation of momentum. 

Statistically, the  momentum transferred to the atom as a cumulative result of such atomic transitions can be written as the difference between the momentum expectation values in the final and the initial state
\begin{align} \label{MomentumShift}
	\Delta P^{\mu}_{\rm atom} &= -\big[\bra{\Psi^{e}_f} \mathds{1}\otimes P^{\mu} \ket{\Psi^{e}_f} - \bra{\Psi^{e}_i} \mathds{1}\otimes P^{\mu} \ket{\Psi^{e}_i}\big],
\end{align}
which can be written, up to second order in the interaction Hamiltonian, in terms of a distribution in the momentum space as
\begin{align}
	\Delta \vb{ P}^{e}_{\rm atom} &\equiv \sum_{\lambda_{\vb{q}}} \int d^3q ~\vb{q} \bra{\Psi^{e}_i} Z(\vb{q}, \lambda_{\vb{q}}) \ket{\Psi^{e}_i},\label{Eq:profile}
\end{align}
with the operator 
\begin{equation}\label{Z}
		Z(\vb{q}, \lambda_{\vb{q}}) \equiv \frac{1}{\hbar} \comm{\int_{-\infty}^{\infty} \dd{\tau'} H_I(\tau')}{\comm{\int_{-\infty}^{\tau'} \dd{\tau} H_I(\tau)}{a^{\dagger}_{\vb{q}\lambda_{\vb{q}}} a^{}_{\vb{q}\lambda_{\vb{q}}}}},
\end{equation}
whose expectation value over all polarizations can be thought to be representing the effective emission profile ${\cal E}_{+}(\vb{q})= \sum_{\lambda_{\vb{q}}}  \bra{\Psi^{e}_i} Z(\vb{q}, \lambda_{\vb{q}}) \ket{\Psi^{e}_i}$ of the atom interacting with the electromagnetic field, i.e.
\begin{eqnarray}
	\Delta \vb{ P}^{e}_{\rm atom} & =  \int d^3q ~\vb{q}  ~{\cal E}_{+}(\vb{q}), \label{MomTransfEms}\\
	\text{while, }
	\mathcal{P}^{e} &=  \int d^3q ~{\cal E}_{+}(\vb{q}),
\end{eqnarray}
marks the total probability of emission. The expression for the force four-vector can be written as
\begin{equation}\label{eq:tensorial_force}
	K^{\mu}  \equiv \frac{\dd{P}^{\mu}}{\dd{\tau}} = \Big(\frac{1}{c}\frac{\dd E}{\dd \tau}, \frac{\dd \mathbf{p}}{\dd \tau}\Big).
\end{equation}
This is the force experienced by the atom  originating from  its interaction with the quantum electromagnetic field. 
The force four-vector in the lab-frame can be obtained by applying Lorentz transformation $\Lambda_\mu^\nu$ on $K^{\mu}$, i.e., $K_{lab}^{\nu} = \Lambda^{\nu}_{\mu} K^{\mu}$. 
A similar force due to absorption can be obtained for the case when the atom is initially in the ground state leading to a final state given by :
\begin{align}
	\ket{\Psi^g_f} = \mathcal{T}\exp\left(-\frac{i}{\hbar}\int_{-\infty}^{\infty} \dd{\tau} H_I(\tau)\right)  \ket{\Psi^{g}_i},
\end{align}
with   $ \ket{\Psi^{g}_i}=\ket{g}\otimes \ket{\Psi}$ and
\begin{eqnarray}
	\Delta \vb{ P}^{a}_{\rm atom}  =  \int d^3q ~\vb{q}  ~{\cal E}_{-}(\vb{q}),  \label{MomTransfAbs}
\end{eqnarray}
being the momentum transfer to the atom from the quantum field environment where  ${\cal E}_{-}(\vb{q}) = - \sum_{\lambda_{\vb{q}}}  \bra{\Psi^{g}_i} Z(\vb{q}, \lambda_{\vb{q}}) \ket{\Psi^{g}_i}$. Consequently, the probability of absorption by the atom is expressed as
\begin{eqnarray}
	\mathcal{P}^{a} &=  \int d^3q ~{\cal E}_{-}(\vb{q}).
\end{eqnarray}
The force experienced by the atom is due to the recoil caused by the stimulated emission from (or absorption by)  the atom.  For  atoms in the  excited state, this force will always be towards the source of light exerting a negative pressure, whereas the force on the atoms in the ground state will be away from the source. In the semi-classical treatment, the absorption and emission probabilities are the same, resulting in equal probabilities of emitting or absorbing a photon which in turn yields force with equal magnitude in the two processes. In a full quantum analysis,  the probabilities of absorption and emission do not remain the same. For a general wavepacket, the absorption and emission probabilities are given by
\begin{align}
	\mathcal{P}^a &= \frac{\pi^3 \abs{\vb{d}}^2}{4\hbar \epsilon_0 c} N h^2(\omega_0,\tilde{k}_0,\sigma)\\
	\mathcal{P}^e &= \frac{\pi^3 \abs{\vb{d}}^2}{4\hbar \epsilon_0 c}    N h^2(\omega_0,\tilde{k}_0,\sigma) +\int_{-\infty}^{\infty} \dd{\tau}~\Gamma_0,
\end{align}
where $ h(\omega_0,\tilde{k}_0,\sigma)$ is a function depending on initial field wavepacket distribution (assumed peaking at $\mathbf{\tilde{k}_{0}}$ with width $\sigma$). The expression for $  h(\omega_0,\tilde{k}_0,\sigma)$ for instance, corresponding to a Gaussian wavepacket is given in Eq.~\eqref{WavePackDist}.

Note that the two probabilities are different and the additional term in the emission probability is due to the spontaneous emission. It is easy to show that despite the probabilities of emission and absorption being different, the forces in these two processes {\it turn out to be exactly the same} in magnitude but opposite in direction,
\begin{eqnarray}
	| \Delta \vb{ P}^{a}_{\rm atom}|=| \Delta \vb{ P}^{e}_{\rm atom}|.
\end{eqnarray}
Therefore, the net force on an ensemble of atoms will depend on the population of atoms in the  excited and  ground states, provided the force expectation is non-zero. In the following, we calculate the force due to $N$-photon Fock state and optical coherent state on an atom due to emission and absorption process.

\section{Atomic emission profile for various field states}\label{Sec:Atomic emission profile for various field states}
In this section we compute $\mathcal{E}_+(\vb{q})$ for the N-photon, coherent, thermal, and the vacuum states of the field.
Using Eq.~\eqref{Z}, the angular emission profile of an excited atom interacting with EM field in state $\ket{\varphi}$ can be evaluated to be
\begin{equation}\label{AngEmiProfile}
	\begin{split}
		&\mathcal{E}_+(\theta_q,\phi_q) = \int_{0}^{\infty} \dd{q} q^2 \mathcal{E}_+(\vb{q}) = \frac{3\Gamma_0}{8\pi} \sin^2\theta_q \int_{- \infty}^{\infty} \dd{\tau} \\
		&+ \frac{\abs{\vb{d}}^2}{\hbar \epsilon_0}  \sum_{\lambda_{\vb{q}}} \int_{0}^{\infty} \dd{q} q^2 
		\int \dd^3k' \sum_{\lambda'} \frac{\sqrt{\omega_{k'}\omega_{q}}}{2(2\pi)^3} \frac{2\pi^2}{c^2}\delta(q-k') \\
		& \times \delta(q-\omega_0/c) \Big\{{\epsilon}^z(\vb{k'},\lambda')  {\epsilon}^z(\vb{q},\lambda_{\vb{q}}) \bra{\varphi} a^{\dagger}_{\vb{k'}\lambda'} a_{\vb{q}\lambda_{\vb{q}}} \ket{\varphi} \\
		& \hspace{1.5cm} +  {\epsilon}^z(\vb{k'},\lambda')  {\epsilon}^z(\vb{q},\lambda_{\vb{q}}) \bra{\varphi}a^{\dagger}_{\vb{q}\lambda_{\vb{q}}} a_{\vb{k'}\lambda'}\ket{\varphi}\Big\}.
	\end{split}
\end{equation}
\subsection{N-photon state}
We first consider the electromagnetic field  to be in $N$-photon wave-packet propagating in the $x$-direction, which is expressed as~\cite{loudon2000quantum}
\begin{equation}\label{wavepacket}
	\ket{N} = \frac{(\mathcal{A}^\dagger_\lambda)^N}{\sqrt{N!}} \ket{0},
\end{equation}
where $\mathcal{A}^\dagger_\lambda = \int d\mathbf{s} ~ \mathcal{F}(\bf s) a^{\dagger}_{\bf s \lambda} $ is the photon-wavepacket creation operator, and $\lambda = 1,2$ denotes the two orthogonal polarizations. The distribution $\mathcal{F}(\vb{s})$ is chosen such that $\abs{\mathcal{F}(\vb{s})}^2$ is a normalized Gaussian~\cite{loudon2000quantum}, 
\begin{align}\label{wavepacketprofile}
	\mathcal{F}(\mathbf{s}) = \left(\frac{1}{2\pi\sigma^2}\right)^{3/4} \exp\left({- \frac{1}{4\sigma^2} (\mathbf{s}-\mathbf{\tilde{k}_{0}})^2}\right),
\end{align}
where  $\mathbf{\tilde{k}_{0}} = \tilde{k}_{0} \hat{\mathbf{x}}$ is the center of the Gaussian wavepacket and  $\sigma$ is the bandwidth. Therefore, the temporal width of the incident pulse is proportional to $(c\sigma)^{-1}$. For pulsed light, width of the Gaussian wave-packet is non-zero, i.e., $\sigma \neq 0$. Since, the electromagnetic field interacting with the atom can cause both, the stimulated emission as well as absorption, we choose $\sigma$  such that the interaction causes not more than one transition in the atom. Using $\bra{N} a^{\dagger}_{\vb{k'}\lambda'} a_{\vb{q}\lambda_{\vb{q}}} \ket{N} = N\mathcal{F}(\vb{q}) \mathcal{F}^*(\vb{k}') \delta_{\lambda' \lambda_0} \delta_{\lambda_0 \lambda_{\vb{q}}}$, Eq.~\eqref{AngEmiProfile} leads to
\begin{equation}\label{AngEmiProfile_N}
	\begin{split}
		&\mathcal{E}_{+}(\theta_{\vb{q}},\phi_{\vb{q}}) = \frac{3\Gamma_0}{8\pi} \sin^2\theta_{\vb{q}} \left(\int_{- \infty}^{\infty} \dd{\tau}\right) \\
		& + N\frac{\pi^3\abs{\vb{d}}^2}{4c\hbar \epsilon_0} \left(\frac{\omega_0}{c}\right)^{5/2} \sin\theta_{\vb{q}} \mathcal{F}(\omega_0/c, \theta_{\vb{q}},\phi_{\vb{q}}) h(\omega_0,\tilde{k}_0,\sigma),
	\end{split}
\end{equation}
where $h(\omega_0,\tilde{k}_0,\sigma)$ is given in Eq.~\eqref{WavePackDist}.
The angular emission rate [see \ref{A:TransitionRates}] resulting from Eq.~\eqref{AngEmiProfile_N} is plotted in Fig.\eqref{fig:EmiProfile} for $N=20$.
\subsection{Coherent state}
Consider the electromagnetic field  to be in a coherent state propagating in the $x$-direction, which is obtained as~\cite{loudon2000quantum}
\begin{equation}\label{coherent}\begin{split}
		\ket{\alpha_\lambda} &= \exp\left(\alpha \mathcal{A}^{\dagger}_{\lambda} -\alpha^* \mathcal{A}_{\lambda} \right) \ket{0}\\
		&= e^{-\frac{1}{2} \abs{\alpha}^2} \sum_{m=0}^{\infty} \frac{\alpha^m }{m!} \left(\mathcal{A}^{\dagger}_{\lambda}\right)^{m}\ket{0},
\end{split}\end{equation}
where $\alpha\in\mathbb{C}$. The mean number of photons  in the state $\ket{\alpha_{\lambda}}$ is $\abs{\alpha}^2 \sum_{\lambda}\int \dd{\vb{s}} \abs{\mathcal{F}(\vb{s})}^2  = \abs{\alpha}^2$. It is straightforward to show that angular emission profile for an excited atom interacting with Em field in state $\ket{\alpha_{\lambda}}$ is given by replacing $N$ in Eq.~\eqref{AngEmiProfile_N} with $\abs{\alpha}^2$.

\subsection{Thermal and the vacuum state}
An electromagnetic thermal bath is described by the density matrix $\rho_{B} = \prod_{\mathbf{k} \lambda} \left(1 - e^{-\beta \hbar \omega_{\bf k}}\right) \exp\left(- \beta \hbar \omega_{\bf k} a^{\dagger}_{\mathbf{k} \lambda} a_{\mathbf{k} \lambda}\right)$ and has the mean number of photons given by $ \mathbb{N}(\omega_q) \equiv \left(e^{\beta \hbar \omega_q} - 1\right)^{-1}$. For an atom immersed in an EM thermal bath, Eq.~\eqref{AngEmiProfile} leads to an angular emission rate given by
\begin{equation}\label{AngEmiProfile_thermal}
	\Gamma_{\text{thermal}}(\theta_{\vb{q}}, \phi_{\vb{q}}) = \frac{3\Gamma_0}{8\pi} \left(\mathbb{N}(\omega_0)+1\right) \sin^2\theta_{\vb{q}}.
\end{equation}
The corresponding expression for the vacuum state of the EM field is obtained by setting $\mathbb{N}(\omega_0) = 0$ in Eq.~\eqref{AngEmiProfile_thermal}. 
\begin{figure}[h!]
	\subfigure[]{
		\includegraphics[width=7cm]{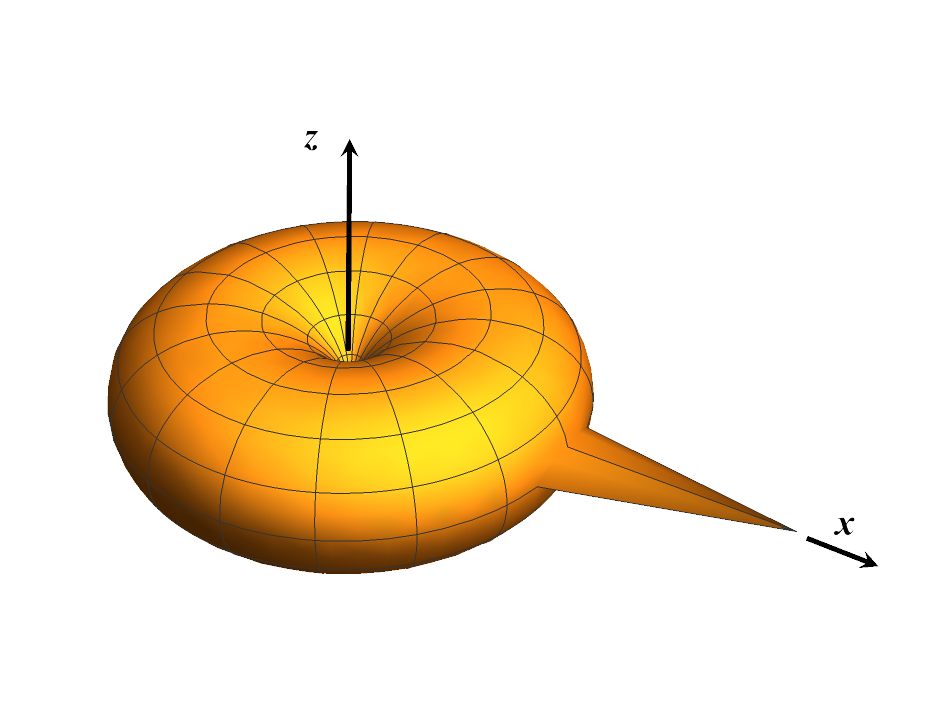}
		\label{fig:StEmi}}
	\subfigure[]{
		\includegraphics[width=7cm]{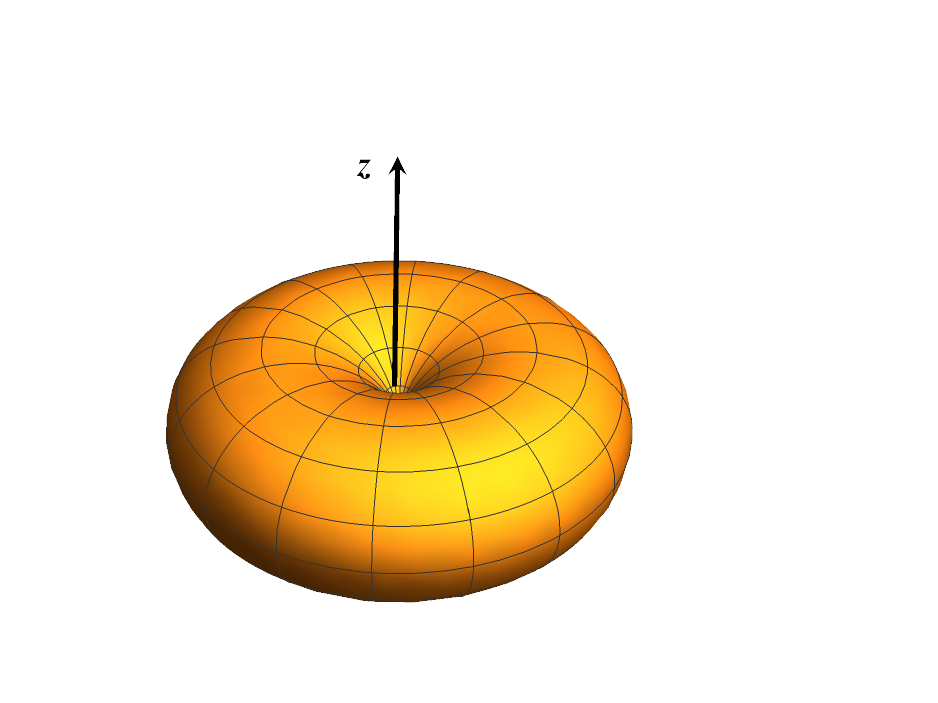}
		\label{fig:SpEmi}}
	\caption{[\ref{fig:StEmi}] Angular emission rate [see Eq.~\eqref{AngEmiProfile_N}] of an atom interacting with an EM $N$-photon pulse traveling along $x$-axis. For the plot we have taken N=20. [\ref{fig:SpEmi}] Angular emission rate for an atom interacting with the thermal and the vacuum states of the EM field. The transition dipole moment $\mathbf{d} = \bra{g}{\mathcal{D}}(\tau=0)\ket{e}$ is assumed to be non-zero only along the $z$-axis.}
	\label{fig:EmiProfile}
\end{figure}
\section{Momentum transfer and the force experienced by the atom}\label{Sec:Momentum transfer and the force experienced by the atom}

\subsection*{$N$-photon and the coherent state}
\begin{figure}
	\subfigure[]{
		\includegraphics[width=7cm]{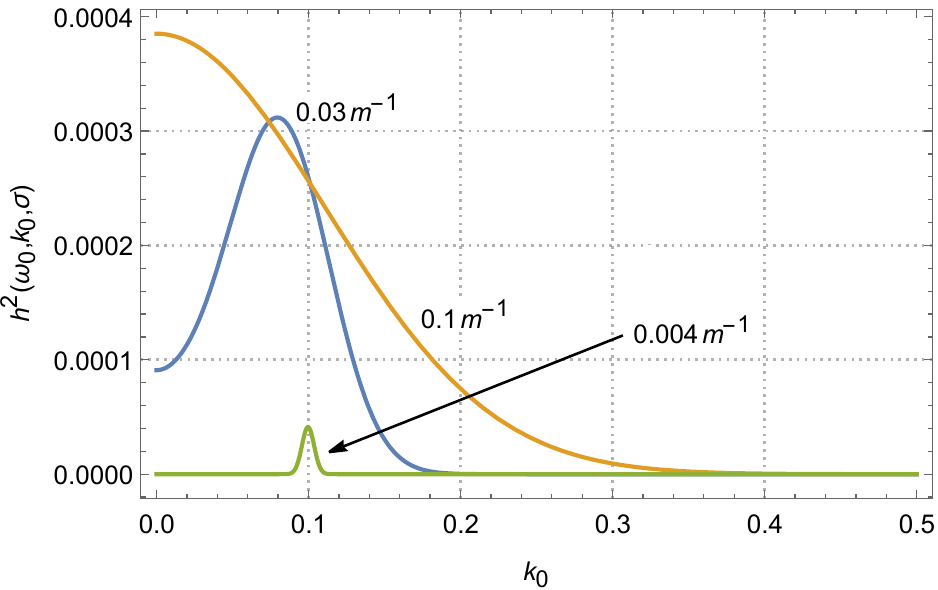}
	\label{h^2_vs_k0}}
	\subfigure[]{
		\includegraphics[width=7cm]{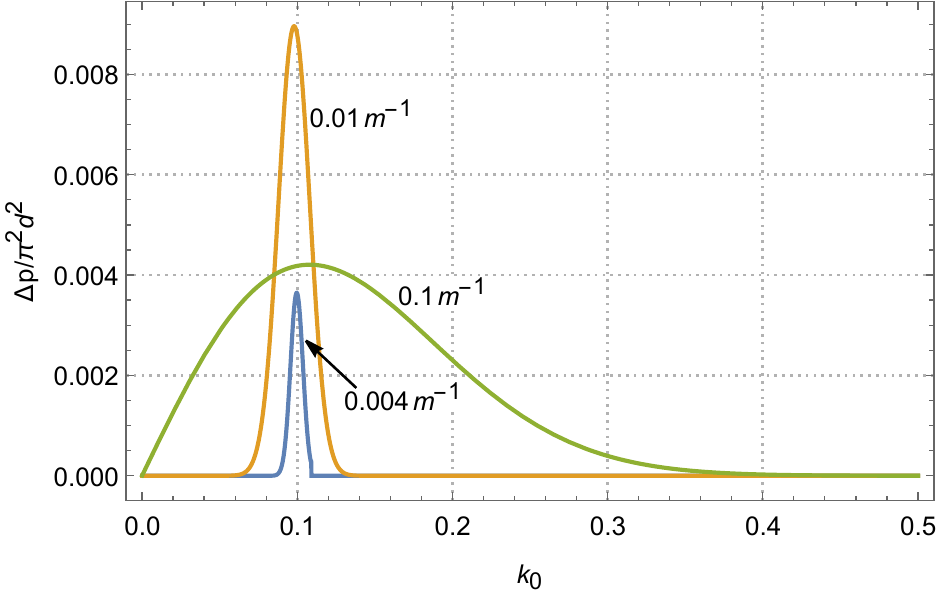}
	\label{p_vs_k0}}
\subfigure[]{
	\includegraphics[width=7cm]{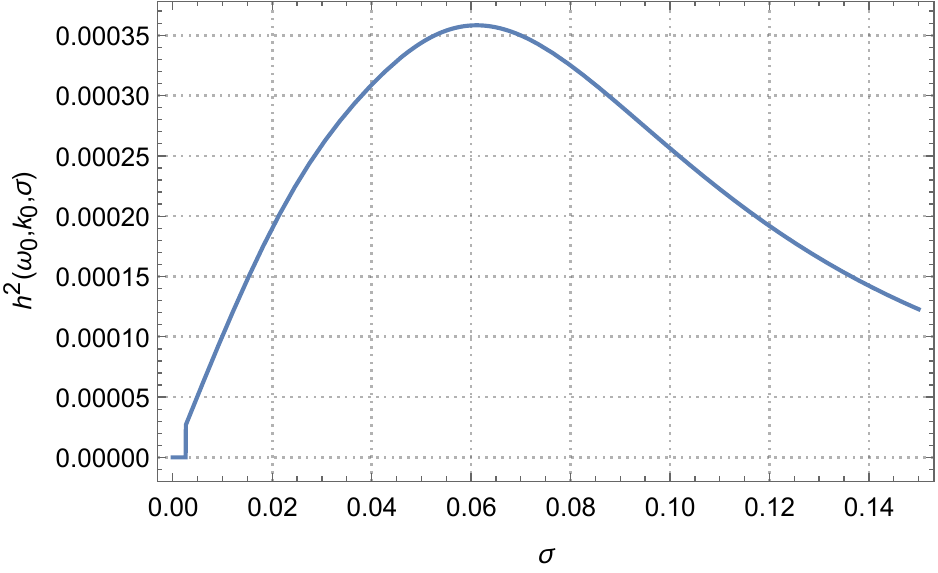}
\label{h^2_vs_sigma}}
\subfigure[]{
	\includegraphics[width=7cm]{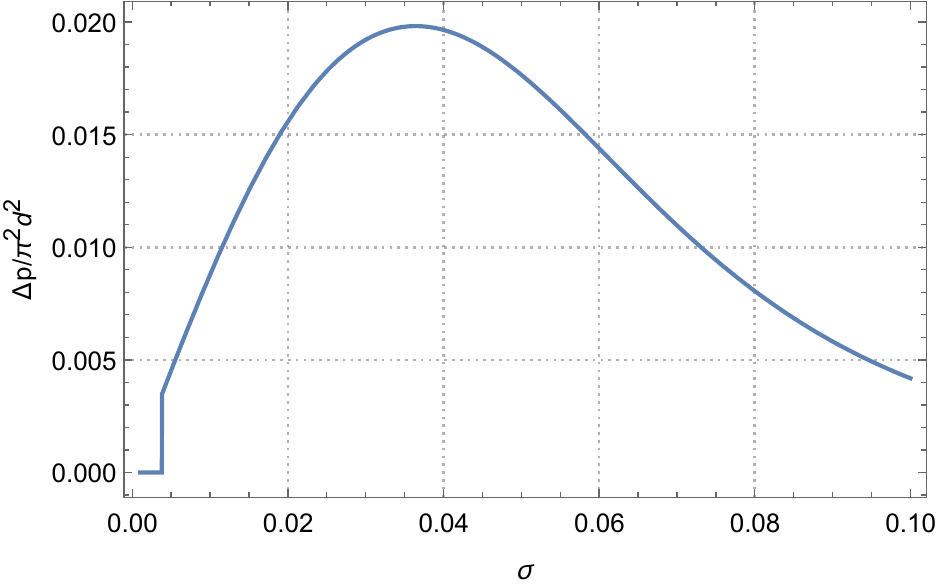}
\label{p_vs_sigma}}
	\caption{Figs.~\ref{h^2_vs_k0} and \ref{p_vs_k0} give the behavior of the atom's stimulated transition probabilities and the momentum transferred to the atom, respectively, as a function of the center $\tilde{k}_0$ of the incident light pulse for different pulse widths $\sigma$. Note that both the transition probabilities and the momentum transferred decrease for increasingly monochromatic (i.e., smaller $\sigma$) incident light pulses. Also note that when $\sigma \approx \omega_0/c$, that is, when the temporal width of the pulse $ \sim (c\sigma)^{-1}$ is of the order of the characteristic time $\sim \omega_0^{-1}$ of the atom, the atom responds appreciably even to far-detuned incident light pulses. Figs.~\ref{h^2_vs_sigma} and \ref{p_vs_sigma} show that the atom becomes transparent to monochromatic incident light, that is, both the transition probabilities and the momentum transferred vanishes as $\sigma \rightarrow 0$. All the plots are for $N=1$.}
	\label{fig:transparency}
\end{figure}

Using Eq.~\eqref{MomentumShift} for the net momentum transfer to the atom due to its interaction with the field, we can obtain the expression for the three-momentum vector to be
\begin{equation}\label{momentum_emission}
	\Delta \vb{P} = - \frac{\pi^2 \abs{\vb{d}}^2}{ 4\pi\epsilon_0} N  h(\omega_0,\tilde{k}_0,\sigma)  h_3(\omega_{0},\tilde{k}_0,\sigma) \vb{x},
\end{equation}
where the expression for $ h_3(\omega_{0},\tilde{k}_0,\sigma)$ is given in Eq.~\eqref{h3}. The momentum transferred as a function of the width $\sigma$, and as a function of $\tilde{k}_0$ for different values of $\sigma$ is plotted in Fig.~\ref{p_vs_k0} and \ref{h^2_vs_sigma} for a single-photon incident pulse. From Fig.~\ref{fig:transparency} we note that contrary to the response of an atom to incident resonant and monochromatic light inside a cavity, an atom in free space becomes transparent to incident monochromatic $(\sigma \rightarrow 0)$ light even if the light pulse is at resonance with the atom (i.e., $\tilde{k}_0 = \omega_0/c$)~\cite{Tjoa2021}.

A useful, as well as simplifying, limit to consider here is the limit 
$\omega_0 \gg c\sigma$. In this limit, the expression for momentum transferred simplifies to (see \ref{A:MomentumTransfer})
\begin{equation}\label{005}
	\begin{split}
		\Delta \vb{P} &=  -\frac{ 2\abs{\vb{d}}^2}{  \sqrt{2\pi}c \epsilon_0} \left(\frac{ \omega_{0}}{ c}\right)^2 N \sigma \vu{x}.
	\end{split}
\end{equation}
The $\omega_0 \gg c\sigma$ regime is of special interest because most of the light pulses available lie in this regime, which makes the results accessible to experimental verification. This regime requires that the incident pulse width be greater than $1/\omega_0 \sim 10^{-14}$s (for optical transitions).
 
Since $c\sigma$ is the spectral width of the incoming pulsed light, the average time $\Delta \tau$ for which the atom and the light interact is proportional to $(c\sigma)^{-1}$. Therefore, the approximate expression for the spatial force reads~\footnote{Alternatively, the force can also be calculated by obtaining transition rate from Eq.~{3} and getting the momentum change per unit time.} 
\begin{equation}\label{force_emission}
	{\bf F} \equiv \frac{\Delta {\bf P}}{\Delta \tau} \equiv \Delta P_x c \sigma {\bf \hat x} \sim  -\frac{ 2\abs{\vb{d}}^2}{  \sqrt{2\pi} \epsilon_0} \left(\frac{ \omega_{0}}{ c}\right)^2 N \sigma^2 \vu{x} .
\end{equation}
Similarly, the average temporal component of the force can be obtained as
\begin{align}
	F^0 \equiv \frac{1}{c} \frac{\Delta E}{\Delta \tau} \sim - \sigma \hbar \omega_0.
\end{align}
Hence, the interaction of the $N$-photon state with an atom initially in the excited state imparts momentum in the direction opposite to the incoming light and causes an attractive force. Further, the magnitude of the force is directly proportional to the photon number $N$. The results for the momentum transfer and the force for the case of coherent state \eqref{coherent} can be obtained simply by replacing $N$ with $\abs{\alpha}^2$ in the corresponding expressions.

Similar calculations for the case when the atom is initially in the ground state can be performed which yields a force which is equal and opposite to the one given in Eq.~\eqref{force_emission}. Therefore, atoms in the ground state experience a repulsive force. Next we calculate the force due to optical coherent states.

From the $N$-photon and optical coherent states of light, it is clear that the emission and absorption processes exert forces which are equal in magnitude and opposite in the direction.

Further, in an atomic ensemble which has a certain distribution of atomic population in the ground and excited states, the signature of the net force will be determined by the population of the two. The force on the ensemble of atoms can be made attractive or repulsive by controlling the excited state populations. 

\subsection*{Vanishing force in the vacuum state of the Electromagnetic field}
Since the electromagnetic field  in the vacuum state can also  be considered as  the coherent state with $\alpha = 0$, the spatial components of the force on the excited as well as on the ground state of a moving atom will be zero. This can be explained as follows: the non-zero momentum transfer to the atom due to its interaction with the field depends upon the expectation value of the $Z(\vb{q},\lambda_{\vb{q}})$ operator [see Eq.~\eqref{Eq:profile}].
This expectation value is non-symmetric for Fock states and optical coherent states [see Fig.~\eqref{fig:StEmi}] which results in non-zero average momentum transfer. However, for isotropic states such as the vacuum (or  thermal) state of the electromagnetic bath this expectation value is symmetric [see Fig.~\eqref{fig:SpEmi}]. Hence Eq.~\eqref{Eq:profile} yields a zero momentum transfer. The extra contribution in the quantum treatment of the emission probabilities is due to the interaction of the atom with the vacuum modes of the environment which yields a symmetric emission profile [see the first term of Eq.~\eqref{AngEmiProfile_N}]. Hence the contribution of this term in the momentum transfer is zero. 
Thus, for vacuum (or the thermal) state of the field, only  the time-like component of the four-force $dE/c\text{d}\tau = {\Gamma_{0} \hbar \omega_{0}}/{c}$ will be non-zero.
The four-force in the lab frame (after Lorentz transformation) reads, 
\begin{align}\label{041}
	K_{lab}^{\nu} = \Lambda^{\nu}_{\mu} K^{\mu} &= \Big(- \gamma \frac{\Gamma_{0} \hbar \omega_{0}}{c}, - \gamma \frac{\Gamma_{0} \hbar \omega_{0}}{c^2} v,0,0\Big),
\end{align}
where $\gamma = (1-v^2/c^2)^{-1/2}$.
An interesting observation from this above expression is that in the lab frame $\frac{\dd \mathbf{p}}{\dd t} = - ({\Gamma_{0} \hbar \omega_{0}}/{c^2} )\mathbf{v}$ is non-zero and the {\it Newtonian} force is proportional to the velocity of the atom. In other words, a lab observer will see a decaying atom  experiencing a ``friction-like'' force~\cite{sonnleitner2017will}. The question now arises is that whether this force as seen from the lab frame has any effect on the motion of the atom. Writing $\vb{p} = \gamma m_0\vb{v}$ we can see easily that the acceleration $\vb{a} \equiv \dd \vb{v}/\dd t$, as seen from the lab  frame 
\begin{equation}\label{accn}
	\mathbf{a} = \frac{1}{\gamma m_{0}}\frac{\dd \mathbf{p}}{\dd t} - \frac{\mathbf{v}}{\gamma m_{0}c^2} \frac{\dd E}{\dd t},
\end{equation}
turns out to be zero. Therefore, the friction-like force acting on the  decaying atom in the lab frame does not cause any acceleration. We now consider the net force on an atomic ensemble which is thermally populated, interacting with a pulsed coherent light.

\subsection*{Force on a thermal atomic ensemble interacting with pulsed coherent light}
We consider an atomic ensemble, containing  $n_0$ number of  mutually non-interacting atoms of mass $m_a$. Initially, the atomic ensemble thermalizes by interacting with a thermal bath at temperature $T$.  This atomic ensemble in the thermal equilibrium is made to interact with a coherent electromagnetic pulse. As a result, we can write an effective rate of change of population of the excited state $n_e$ and the ground state $n_g$ as $\dot{n}_e =-\Gamma_{e\rightarrow g} n_e + \Gamma_{g\rightarrow e} n_g$, where $\Gamma_{i\to j}$ is the rate of transition from state $\ket{i}$ to $\ket{j}$.

If the system attains an equilibrium configuration, then the number of upward transitions will be equal to the number of downward transitions. Since, the force in the two transitions is equal in magnitude but opposite in sign, the net force in the new equilibrium can be written as
\begin{align}
	\mathbf{F}_{net} = \mathbf{F} (n_g-n_e ) \equiv \mathbf{F} n_0(1-2\chi ),\label{Eq:ThermalForce},
\end{align}
where $\mathbf{F}$ is the force acting on an atom in the ground state interacting with a pulsed coherent state. The $\chi = n_e/n_0$ is the fraction of the excited population. From here it is clear that when the fraction of excited atoms is more than half, the direction of the force vector changes. In an initial thermal distribution, the temperature $T$ and mean number of photons $\abs{\alpha}^2$ in the coherent state  decide the fraction $\chi$ in the new equilibrium and is given by
\begin{equation}
	\chi = \frac{\Gamma_{0} \mathbb{N}(\omega_{0})+\Gamma^a}{\Gamma_{0} \coth\Big(\frac{\beta \hbar \omega_{0}}{2}\Big) + \tilde{\Gamma}^e +\Gamma^a},\label{Eq:chi}
\end{equation}
where $\beta = 1/k_B T$ and we have used $\Gamma_{g\rightarrow e} = \Gamma_{0} \mathbb{N}(\omega_{0})+\Gamma^a$, $\Gamma_{e\rightarrow g} = \Gamma_{0}\mathbb{N}(\omega_{0}) + \tilde{\Gamma}^e + \Gamma_0$, with
\begin{subequations}\label{transition_rates}
	\begin{equation}
		\Gamma^e \sim \frac{ \pi^3 \abs{\vb{d}}^2}{4\hbar \epsilon_0 }    \abs{\alpha}^2 \sigma h^2(\omega_0,\tilde{k}_0,\sigma) + \Gamma_0 \equiv \tilde{\Gamma}^e + \Gamma_{0},
	\end{equation}
	\begin{equation}
		\Gamma^a \sim \frac{ \pi^3 \abs{\vb{d}}^2}{4\hbar \epsilon_0 }    \abs{\alpha}^2 \sigma h^2(\omega_0,\tilde{k}_0,\sigma).
	\end{equation}
\end{subequations} 
Note that $\Gamma_0 \mathbb{N}(\omega_0)$ comes from the atom's interaction with the thermal bath of photons, $\tilde{\Gamma}^e$ and $\Gamma^a$ contributions come from the atom's interaction with the coherent state, and $\Gamma_0$ is the spontaneous decay rate of the atom in vacuum~(see \ref{A:TransitionRates}).
Combining Eqs.~\eqref{Eq:ThermalForce} and~\eqref{Eq:chi} yields the exact force on an ensemble of atoms due to pulsed coherent states.

\section{Discussion and Conclusions}\label{Sec:Discussion and Conclusions}
In this paper, we have analyzed the interaction of quantum electromagnetic field with a moving two-level atom. Through the interaction via the dipole the atom undertakes absorption or emission. In this process the atom gets a push or a pull in order to conserve the momentum. For a classical electromagnetic field, the rate of emission is equal to the rate of absorption, leading to the same magnitude of the resultant force, but opposite in direction. However, as we have seen above, the quantum treatment of the electromagnetic field results in unequal rates of emission and absorption, but still the magnitude of the force of recoil remains the same for the two processes. The emission part, apart from the stimulated emission, comprises of the spontaneous emission as well, which originates from the vacuum structure of the quantum field. However, this extra part of  emission  does not lead to any extra recoil as the effective emission from the spontaneous emission is isotropic. This further establishes that the vacuum state of the field does not lead to any pushing or pulling force. In fact, the Eqs. (\ref{MomTransfEms},\ref{MomTransfAbs}) suggest that for the states which lead to ${\cal E}_{\pm}({\bf q})$ being  symmetric functions of ${\bf q}$, the momentum transfer to/from the atom vanishes. For example, for a thermal state of the field the net momentum transfer is zero. On the other hand, for the states  of a certain class which do not give rise to symmetric  ${\cal E}_{\pm}({\bf q})$, e.g., the $N$-photon state or the coherent state, there is a net momentum transfer leading to a non-zero force on an  atom.

For an atom in the ground state the force is pushing while for the atom in the excited state the force is pulling, which turns out  to be equal in magnitude to the pushing force in absorption. Thus, in an ensemble of atoms the net force on the system is proportional to the difference in the number of atoms in the ground state and those in the excited state $(n_g-n_e)$.  Interestingly, therefore, on a system having population inversion $n_e>n_g$, the ensemble experiences a net pulling force, i.e. a net negative pressure rather than a pushing force of standard radiation pressure. This analysis also demystifies the appearance of a drag force as reported in~\cite{sonnleitner2017will} and identifies the class of states which gives rise to a net force and which do not. One of the interesting application of this analysis could be to trap and/or isolate atoms with certain velocity in an atomic ensemble. This provides an additional control for optical tweezers and laser cooling techniques.

\section*{Acknowledgments}
	NA acknowledges financial support from the University Grants Commission (UGC), Government of India, in the form of a research fellowship (Sr.~No.~2061651285). Research of KL is partially supported by the Start-up Research Grant of SERB, Government of India (SRG/2019 /002202). SKG acknowledges  the financial support from SERB-DST (File No. ECR/2017/002404). NB wishes to thank IISER Mohali for hospitality during this work.

\appendix
\section{Transition Probability}\label{A:TransitionP}
In the following, the atom-field states are denoted by $\ket{\Psi}$ and $ \ket{\phi} \in \{\ket{g},\ket{e}\} $ denotes the atomic states. $\ket{\bar{\phi}}$ denotes the ``complement'' state of $\ket{\phi}$, for example, if $\ket{\phi} = \ket{g}$, then $\ket{\bar{\phi}} = \ket{e}$.
We aim to write the general expression for the transition probability. If the initial atomic state is $\ket{\phi_i}$, then
\begin{equation*}
	\begin{split}
	1 &= \abs{\braket{\bar{\phi}_i}{\Psi_f}}^2 + \abs{\braket{{\phi}_i}{\Psi_f}}^2 \\
	&= \braket{\Psi_f}{\bar{\phi}_i}\braket{\bar{\phi}_i}{\Psi_f} + \braket{\Psi_f}{{\phi}_i}\braket{{\phi}_i}{\Psi_f}\\
	&=  \bra{\Psi_f}\bar{\mathbb{P}}_i \ket{\Psi_f} + \bra{\Psi_f}{\mathbb{P}}_i \ket{\Psi_f},
	\end{split}
\end{equation*}
where $\mathbb{P}_i \equiv \dyad{\phi_i}{\phi_i}$ is the projector on the atomic state $\ket{\phi_i}$. For a two-level atom $\mathbb{P}_i + \bar{\mathbb{P}}_i = 1$.
The transition probability, $\mathcal{P} = \bra{\Psi_f}\bar{\mathbb{P}}_i \ket{\Psi_f}$, can be cast into a form containing only the initial states and the time-evolution operator using $\ket{\Psi_f} = U \ket{\Psi_i}$, where $U = \mathcal{T}\exp\left(-\frac{i}{\hbar}\int_{-\infty}^{\infty} \dd{\tau} H_I(\tau)\right)$ is the unitary time evolution operator, as
\begin{equation}
	\mathcal{P} = \bra{\Psi_f}\bar{\mathbb{P}}_i \ket{\Psi_f} = \bra{\Psi_i} U^{\dagger}\bar{\mathbb{P}}_i U\ket{\Psi_i} = \bra{\Psi_i} U^{\dagger} (1 - {\mathbb{P}}_i) U\ket{\Psi_i}.
\end{equation}
Further, using 
	\begin{equation}
		\begin{split}
			&U^{\dagger} (1 - {\mathbb{P}}_i) U = (1 - {\mathbb{P}}_i) + \frac{i}{\hbar} \comm{\int_{- \infty}^{\infty} \dd{\tau} H_I(\tau)}{ (1 - {\mathbb{P}}_i)} \\
			&- \frac{1}{\hbar^2} \comm{\int_{- \infty}^{\infty} \dd{\tau'} H_I(\tau')}{\comm{ \int_{- \infty}^{\tau'} \dd{\tau} H_I(\tau)}{ (1 - {\mathbb{P}}_i)}} + \cdots,
		\end{split}
	\end{equation}
to second order in the interaction Hamiltonian, the transition probability takes the form
\begin{multline*}
	\mathcal{P} = \bra{\Psi_i} (1 - {\mathbb{P}}_i)\ket{\Psi_i} + \bra{\Psi_i}\frac{i}{\hbar} \comm{\int_{-\infty}^{\infty} \dd{\tau} H_I(\tau)}{(1 - {\mathbb{P}}_i)}\ket{\Psi_i} \\ + \bra{\Psi_i}  \left(\frac{i}{\hbar}\right)^2 \comm{\int_{-\infty}^{\infty} \dd{\tau'} H_I(\tau')}{\comm{\int_{-\infty}^{\tau'} \dd{\tau} H_I(\tau)}{(1 - {\mathbb{P}}_i)}} \ket{\Psi_i}.
\end{multline*}
First and the second terms vanish, leaving us with
\begin{equation}\label{TransP}
	\begin{split}
		\mathcal{P} = \bra{\Psi_i} Z_1 \ket{\Psi_i},
	\end{split}
\end{equation} where
\begin{equation}\label{Z}
	Z \equiv \left(\frac{i}{\hbar}\right)^2 \comm{\int_{-\infty}^{\infty} \dd{\tau'} H_I(\tau')}{\comm{\int_{-\infty}^{\tau'} \dd{\tau} H_I(\tau)}{(1 - {\mathbb{P}}_i)}}.
\end{equation}

\section{Momentum Transfer}\label{A:MomentumTransfer}
Net momentum transferred to the atom is given by 
\begin{equation}\label{MomentumChange}
	\begin{split}
	-\left(\vb{P}_f - \vb{P}_i \right) &= -\hbar \sum_{\lambda_{\vb{q}}} \int d^3q ~\vb{q} \bra{\Psi_f} a^{\dagger}_{\vb{q}\lambda_{\vb{q}}} a^{}_{\vb{q}\lambda_{\vb{q}}}\ket{\Psi_f} \\
	& + \hbar \sum_{\lambda_{\vb{q}}}  \int d^3q ~\vb{q} \bra{\Psi_i} a^{\dagger}_{\vb{q}\lambda_{\vb{q}}} a^{}_{\vb{q}\lambda_{\vb{q}}}\ket{\Psi_i},
	\end{split}
\end{equation} 
which can be written as
\begin{equation}
	\begin{split}
	-\left(\vb{P}_f - \vb{P}_i \right) &= -\hbar \sum_{\lambda_{\vb{q}}} \int d^3q ~\vb{q} \bra{\Psi_i} U^{\dagger} a^{\dagger}_{\vb{q}\lambda_{\vb{q}}} a^{}_{\vb{q}\lambda_{\vb{q}}} U \ket{\Psi_i} \\
	& + \hbar \sum_{\lambda_{\vb{q}}}  \int d^3q ~\vb{q} \bra{\Psi_i} a^{\dagger}_{\vb{q}\lambda_{\vb{q}}} a^{}_{\vb{q}\lambda_{\vb{q}}}\ket{\Psi_i}.
	\end{split}
\end{equation}
Using
\begin{equation}
	\begin{split}
		&U^{\dagger} a^{\dagger}_{\vb{q}\lambda_{\vb{q}}} a^{}_{\vb{q}\lambda_{\vb{q}}} U = a^{\dagger}_{\vb{q}\lambda_{\vb{q}}} a^{}_{\vb{q}\lambda_{\vb{q}}} + \frac{i}{\hbar} \comm{\int_{- \infty}^{\infty} \dd{\tau} H_I(\tau)}{ a^{\dagger}_{\vb{q}\lambda_{\vb{q}}} a^{}_{\vb{q}\lambda_{\vb{q}}}} \\
		&- \frac{1}{\hbar^2} \comm{\int_{- \infty}^{\infty} \dd{\tau'} H_I(\tau')}{\comm{ \int_{- \infty}^{\tau'} \dd{\tau} H_I(\tau)}{ a^{\dagger}_{\vb{q}\lambda_{\vb{q}}} a^{}_{\vb{q}\lambda_{\vb{q}}}}} + \cdots,
	\end{split}
\end{equation}
to second order in the interaction Hamiltonian, we get
\begin{equation}
	\begin{split}
	&\left(\vb{P}_f - \vb{P}_i \right) \\
	&= \hbar \sum_{\lambda_{\vb{q}}} \int d^3q ~\vb{q} \bra{\Psi_i} \frac{i}{\hbar} \comm{\int_{-\infty}^{\infty} \dd{\tau} H_I(\tau)}{a^{\dagger}_{\vb{q}\lambda_{\vb{q}}} a^{}_{\vb{q}\lambda_{\vb{q}}}} \ket{\Psi_i} \\
	&+ \hbar \sum_{\lambda_{\vb{q}}} \int d^3q ~\vb{q} \\
	& \times \bra{\Psi_i} \left(\frac{i}{\hbar}\right)^2 \comm{\int_{-\infty}^{\infty} \dd{\tau} H_I(\tau)}{\comm{\int_{-\infty}^{\tau} \dd{\tau'} H_I(\tau')}{a^{\dagger}_{\vb{q}\lambda_{\vb{q}}} a^{}_{\vb{q}\lambda_{\vb{q}}}}} \ket{\Psi_i}.
	\end{split}
\end{equation}
Further, $\bra{\Psi_i} \frac{i}{\hbar} \comm{\int \dd{\tau} H_I(\tau)}{a^{\dagger}_{\vb{q}\lambda_{\vb{q}}} a^{}_{\vb{q}\lambda_{\vb{q}}}} \ket{\Psi_i} = 0$ leads to 
\begin{multline}
	\begin{split}
	&-\left(\vb{P}_f - \vb{P}_i \right) = -\hbar\sum_{\lambda_{\vb{q}}} \int d^3q ~\vb{q} \left(\frac{i}{\hbar}\right)^2 \\
	& \times \bra{\Psi_i}   \comm{\int_{-\infty}^{\infty} \dd{\tau} H_I(\tau)}{\comm{\int_{-\infty}^{\tau} \dd{\tau'} H_I(\tau')}{a^{\dagger}_{\vb{q}\lambda_{\vb{q}}} a^{}_{\vb{q}\lambda_{\vb{q}}}}} \ket{\Psi_i}.
	\end{split}
\end{multline}
That is, \begin{equation}\label{momentum_f-i}
	\left(\vb{P}_f - \vb{P}_i \right)= -\sum_{\lambda_{\vb{q}}} \int d^3q ~\vb{q} \bra{\Psi_i} Z_{\vb{q}\lambda_{\vb{q}}} \ket{\Psi_i},
\end{equation} where \begin{equation}\label{Zq}
	Z(\vb{q},\lambda_{\vb{q}}) \equiv \frac{1}{\hbar}  \comm{\int_{-\infty}^{\infty} \dd{\tau} H_I(\tau)}{\comm{\int_{-\infty}^{\tau} \dd{\tau'} H_I(\tau')}{a^{\dagger}_{\vb{q}\lambda_{\vb{q}}} a^{}_{\vb{q}\lambda_{\vb{q}}}}}.
\end{equation}
It is further easy to establish that the probability of transition in Eq.(\ref{TransP}) can be expressed as
\begin{equation}
	\begin{split}
	\mathcal{P} &= \bra{\Psi_i} Z \ket{\Psi_i} \\
	&=  \pm \int \dd^3{q} \sum_{\lambda_{\vb{q}}}\bra{\Psi_i} Z(\vb{q}, \lambda_{\vb{q}})\ket{\Psi_i} \equiv \int \dd^3{q} ~ {\cal E}_{\pm}(\vb{q}), 
	\end{split}
\end{equation}
with $\pm$ signifying process of emission and absorption respectively. It is easy to show that ${\cal E}_{\pm}(\vb{q})$ is non-negative for all $\vb{q}$.

\section{Calculations for number state}\label{A:NumberState}

Using $\ket{\Psi_f} = U \ket{\Psi_i}$, the final state of the atom-field composite system for the emission process, to order $\abs{\vb{d}}^2$ in the interaction Hamiltonian, can be written as
\begin{equation}\label{emi_final_state}
	\begin{split}
	&\ket{\Psi^e_f} = \ket{e} \otimes \ket{N} \left\{1 - \frac{\Gamma_0}{2}  \int_{- \infty}^{\infty} \dd{\tau} \right\} \\
	&+  \frac{ 2\pi \vb{d}}{\sqrt{\hbar \epsilon_0}} \ket{g}   \otimes   \int \frac{ d^3k~k}{\sqrt{2 \omega_{k} (2\pi)^3}}  \sum_{\lambda=1}^{2}  {\mathbf{\epsilon}}^z({\bf k},\lambda) \delta\left(k-\frac{\omega_0}{c}\right) a^{\dagger}_{{\bf k}\lambda}   \ket{N} \\
	&- \sqrt{N}\frac{\abs{\vb{d}}^2}{\hbar \epsilon_0} \ket{e} \otimes  \int \frac{  d^3k'}{\sqrt{2 \omega_{k'} (2\pi)^3}}  \int \frac{  d^3k}{\sqrt{2 \omega_{k} (2\pi)^3}} \\
	& \times \sum_{\lambda=1}^{2} \omega_{k}  \epsilon^z({\bf k},\lambda) \epsilon^z({\bf k'},\lambda_0) g(k,k')  \mathcal{F}(\vb{k}') a^{\dagger}_{{\bf k}\lambda}\ket{N-1}
	\end{split}
\end{equation}
 where $\Gamma_0 = \abs{\vb{d}}^2 \omega_0^3/(3\pi\hbar \epsilon_0 c^3)$ is the atomic spontaneous decay rate in vacuum, 
\begin{equation} \label{WavePackDist}
	\begin{split}
	h(\omega_{0},\tilde{k}_{0},\sigma) &=  \Big(\frac{1}{2\pi\sigma^2}\Big)^{3/4}   \Big(\frac{\omega_{0}}{c}\Big)^{5/2} e^{- \frac{1}{4\sigma^2} [\frac{\omega_{0}^2}{c^2} + \tilde{k}_{0}^2]} \\
	&\times  \Big[I_{0}^{2}\Big(\frac{\omega_{0} \tilde{k}_{0}}{4 c \sigma^2}\Big) + I_{1}^{2}\Big(\frac{\omega_{0} \tilde{k}_{0}}{4 c \sigma^2}\Big)\Big],
	\end{split}
\end{equation} 
and $I_{\nu}(x)$ is the modified Bessel function of order $\nu$ and $I^2_{\nu}(x)$ means $\left(I_{\nu}(x)\right)^2$.
Note that whenever we deal with terms like $(\omega_0-kc)^{-1}$, using~\cite{tannoudji1992atom}
\begin{equation}\label{lamb}
	\lim_{\epsilon \rightarrow 0^+} \frac{1}{x \pm i \epsilon} = \text{P.V.}\left(\frac{1}{x}\right) \mp i\pi \delta(x),
\end{equation}
in the integrals, we retain only the second term in Eq.~\eqref{lamb}. This is because the first term corresponds to the energy shift of the atomic levels and the second term describes the dissipative phenomenon introduced by the interaction of the atom with the EM field~\cite{tannoudji1992atom}. We assume that the energy shift of the atomic levels has already been incorporated in $\omega_0$.

From (\ref{emi_final_state}), we obtain the emission probability to be
\begin{equation}\label{emission_prob}
	\mathcal{P}^e = \frac{ \pi^3 \abs{\vb{d}}^2}{4\hbar \epsilon_0 c}    N h^2(\omega_0,\tilde{k}_0,\sigma) + \Gamma_0 \int_{-\infty}^{\infty}\dd{\tau}.
\end{equation}
Using (\ref{MomentumChange}), the expression for the net momentum transferred to the atom, due to atom-field interactions up to order $\abs{\vb{d}}^2$, comes out to be
\begin{equation}
	\Delta \vb{P} = - \frac{\pi^2 \abs{\vb{d}}^2}{ 4\pi\epsilon_0} N  h(\omega_0,\tilde{k}_0,\sigma)  h_3(\omega_{0},\tilde{k}_0,\sigma) \vb{x},
\end{equation}
where
\begin{equation}\label{h3}
	\begin{split}
	&h_3(\omega_0, \tilde{k}_0, \sigma) \equiv  \frac{\sqrt[4]{2} \pi ^{5/4} \omega_0 ^{5/2} }{c^{9/2} \tilde{k}_0  \sigma ^{3/2}} e^{-\frac{c^2 \tilde{k}_0 ^2+\omega_0 ^2}{4 c^2 \sigma ^2}} \\
	& \times \Bigg[ I_1\left(\frac{\tilde{k}_0  \omega_0 }{4 c \sigma ^2}\right) \left(\tilde{k}_0  \omega_0  I_0\left(\frac{\tilde{k}_0  \omega_0 }{4 c \sigma ^2}\right)-2 c \sigma ^2 I_1\left(\frac{\tilde{k}_0  \omega_0 }{4 c \sigma ^2}\right)\right)\Bigg].
	\end{split}
\end{equation}
In the monochromatic wavepacket limit, that is $\omega_{0}\gg c\sigma$, using the asymptotic form~\cite{Arfken2013}
\begin{equation}
	I^{n}_{\nu}\Big(\frac{x}{n}\Big) \rightarrow \frac{e^{x}}{(2 \pi)^{n/2} (\frac{x}{n})^{n/2}},
\end{equation} of the modified Bessel functions, various expressions containing $h(\omega_{0},\tilde{k}_{0},\sigma)$ and $h_3(\omega_{0},\tilde{k}_{0},\sigma)$ can be simplified. For $\tilde{k}_0 = \omega_{0}/c$,  in the $\omega_{0}\gg c\sigma$ limit we obtain \begin{align}
	h^2(\omega_{0},\tilde{k}_{0},\sigma) &\to \Big(\frac{1}{2\pi}\Big)^{3/2}   \Big(\frac{\omega_{0}}{c}\Big) \frac{16\sigma}{\pi^2},~\text{and}\\
	h(\omega_{0},\tilde{k}_{0},\sigma) &h_3(\omega_{0},\tilde{k}_{0},\sigma) \to  \frac{8}{\pi \sqrt{2\pi}c}  \left(\frac{ \omega_{0}}{ c}\right)^2  \sigma,
\end{align}
which lead to, for example, an emission probability given by
\begin{equation}
	\mathcal{P}^e =  \frac{2 \abs{\vb{d}}^2}{\sqrt{2\pi}\hbar \epsilon_0 c}  \Big(\frac{\omega_{0}}{c}\Big) N  \sigma + \Gamma_0 \int_{-\infty}^{\infty}\dd{\tau}.
\end{equation}
Similarly, we can obtain the Eq.~(18) of the paper.
\section{Transition Rates}\label{A:TransitionRates}
From Eq.~\eqref{emission_prob}, we note that $\mathcal{P}^e \equiv \mathcal{P}^e_{\text{st}} + \mathcal{P}^e_{\text{sp}}$, has two contributions, $\mathcal{P}^e_{\text{st}}$ comes from the stimulated emission and persists for a time interval proportional to the temporal width $\Delta t_{\text{st}} \sim (c\sigma)^{-1}$ of the incident light pulse, and $\mathcal{P}^e_{\text{sp}}$ comes from the spontaneous decay of the atom and persists for a time interval $\Delta t_{\text{sp}} = \int_{- \infty}^{\infty} \dd{\tau}$. Using these observations, the corresponding decay rates can be computed to be $\Gamma^e_{\text{st}} \equiv \mathcal{P}^e_{\text{st}}/\Delta t_{\text{st}}$ and $\Gamma_{0} \equiv \mathcal{P}^e_{\text{sp}}/\Delta t_{\text{sp}}$, respectively. The total decay rate, $\Gamma^e = \Gamma^e_{\text{st}} + \Gamma^e_{\text{sp}}$, comes out to be
\begin{equation}
	\Gamma^e \sim \frac{ \pi^3 \abs{\vb{d}}^2}{4\hbar \epsilon_0 }    N \sigma h^2(\omega_0,\tilde{k}_0,\sigma) + \Gamma_0.
\end{equation}
Similarly,
\begin{equation}
	\Gamma^a \sim \frac{ \pi^3 \abs{\vb{d}}^2}{4\hbar \epsilon_0 }    N \sigma h^2(\omega_0,\tilde{k}_0,\sigma).
\end{equation}
The corresponding expressions, Eq.~(26) in the paper, for the atom's interaction with the coherent state are obtained by replacing $N$ with $\abs{\alpha}^2$.

\end{document}